# COMPUTATIONAL COMPLEXITY COMPARISON OF MULTI-SENSOR SINGLE TARGET DATA FUSION METHODS BY MATLAB

Sayed Amir Hoseini[1] and Mohammad Reza Ashraf [2]

[1]Department of Electrical Engineering, Amirkabir University of technology, Tehran, Iran
`a.hoseini@aut.ac.ir`
[2] Department of Electrical Engineering, University of Tehran, Tehran, Iran
`m.r.ashraf@ut.ac.ir`

*ABSTRACT*

*Target tracking using observations from multiple sensors can achieve better estimation performance than a single sensor. The most famous estimation tool in target tracking is Kalman filter. There are several mathematical approaches to combine the observations of multiple sensors by use of Kalman filter. An important issue in applying a proper approach is computational complexity. In this paper, four data fusion algorithms based on Kalman filter are considered including three centralized and one decentralized methods. Using MATLAB, computational loads of these methods are compared while number of sensors increases. The results show that inverse covariance method has the best computational performance if the number of sensors is above 20. For a smaller number of sensors, other methods, especially group sensors, are more appropriate..*

*KEYWORDS*

*Data fusion, Target Tracking, Kalman Filter, Multi-sensor, MATLAB*

## 1. INTRODUCTION

Data fusion is the process of combining information from a number of different sources to provide a robust and complete description of an environment or process of interest. Data fusion is of special significance in any application where large amounts of data must be combined, fused and distilled to obtain information of appropriate quality and integrity on which decisions can be made. Data fusion finds application in many military systems, in civilian surveillance and monitoring tasks, in process control and in information systems. Data fusion methods are particularly important in the drive toward autonomous systems in all these applications [1].

Estimation is the single most important problem in sensor data fusion. Fundamentally, an estimator is a decision rule which takes as an argument a sequence of observations and whose action is to compute a value for the parameter or state of interest. Almost all data fusion problems involve this estimation process: we obtain a number of observations from a group of sensors and using this information we wish to find some estimate of the true state of the environment we are observing. Estimation encompasses all important aspects of the data fusion problem. The most famous estimation tool in target tracking is Kalman filter.

Tracking targets with Kalman filtering is an active research area and there are substantial literatures in this field such as [2], [3] and [4]. [1,4] talk about several multi-sensor data fusion methods. [5] and [6] are also the same but latest attempts. In [1], the computational complexities of these methods are compared based on mathematical formulation, but no simulation or implementation proof is presented. In this paper, the aim is to continue the related work in [1] and evaluate its claim using MATLAB simulations.

This paper begins with a brief summary of the Kalman filter algorithm. The intention is to introduce notation and key data fusion concepts; Prior familiarity with the basic Kalman Filter algorithm is assumed. The multi-sensor Kalman filter is then discussed. Four main algorithms are considered; the group-sensor method, the sequential sensor method, the inverse covariance form and the track-to-track fusion. For the related concepts and additional formulation refer to [1]. After that, in section 3, these methods are evaluated in theory and simulation. Finally, Simulation results and a comparison of these methods are presented in section 4.

## 2. KALMAN FILTER IN TARGET TRACKING

The Kalman Filter is a recursive linear estimator which successively calculates an estimate for a continuous valued state, that evolves over time, on the basis of periodic observations that of this state.

### 2.1. State and Observation Models

The starting point for the Kalman filter algorithm is to define a model for the states to be estimated in the standard state-space form and observation model for data received from sensor [7]. A general motion model used in discrete Kalman filter for target tracking is:

$$x(k) = F(k)x(k-1) + G(k)v(k) \qquad (1)$$

$$z(k) = H(k)x(k) + w(k) \qquad (2)$$

where F(k) is the state transition matrix, x(k) is the state vector at time k, G(k) is the process noise gain matrix, The process noise v(k) and the measurement noise w(k) are zero mean, mutually independent, white, Gaussian with covariance Q and R respectively. z(k) is the measurement vector at time k and H(k) is observation matrix of the states computed at time k.

$$E\{v(k)\} = E\{w(k)\} = 0, \quad \forall k, \qquad (3)$$

$$E\{v(i)v^T(j)\} = \delta_{ij}Q(i), \quad E\{w(i)w^T(j)\} = \delta_{ij}R(i). \qquad (4)$$

#### 2.1.1. State Prediction

The state and state covariance matrix at time k-1 are predicted to time k as follows:

$$\hat{x}(k|k-1) = F(k)\hat{x}(k-1|k-1) \qquad (5)$$

$$P(k|k-1) = F(k)P(k-1|k-1)F^T(k) + G(k)Q(k)G^T(k), \qquad (6)$$

where $\hat{x}(k-1|k-1)$ is the estimated state vector at time $k$, $P(k-1|k-1)$ is the estimated state covariance matrix at the same time, $\hat{x}(k|k-1)$ is the predicted state and $P(k|k-1)$ is the predicted state covariance matrix.

#### 2.1.2. Measurement update

At time k an observation $z(k)$ is made and the updated estimate $\hat{x}(k|k)$ of the state $x(k)$, together with the updated estimate covariance $P(k|k)$ is computed from the state prediction and observation according to:

$$\hat{x}(k|k) = \hat{x}(k|k-1) + W(k)(z(k) - H(k)\hat{x}(k|k-1)) \qquad (7)$$

and

$$P(k|k) = (1 - W(k)H(k))P(k|k-1)(1 - W(k)H(k))^T + W(k)R(k)W^T(k) \qquad (8)$$

where the gain matrix $W(k)$ is given by:

$$W(k) = P(k|k-1)H(k)\left[H(k)P(k|k-1)H^T(k) + R(k)\right]^{-1} \tag{9}$$

## 2.2. The Multi-Sensor Kalman Filter

Many of the techniques developed for single sensor Kalman filters can be applied directly to multi-sensor estimation and tracking problems. In principle, a group of sensors can be considered as a single sensor with a large and possibly complex observation model. In this case the Kalman filter algorithm is directly applicable to the multi-sensor estimation problem. This technique is called "group-sensor method". However, as will be seen, this approach is practically limited to relatively small numbers of sensors.

A second approach is to consider each observation made by each sensor as a separate and independent realization, made according to a specific observation model, which can be incorporate into the estimate in a sequential manner. Again, single-sensor estimation techniques, applied sequentially, can be applied to this formulation of the multi-sensor estimation problem. This technique is called "sequential-sensor method". However, as will be seen, this approach requires that a new prediction and gain matrix be calculated for each observation from each sensor at every time-step, and so is computationally very expensive.

A third approach is to explicitly derive equations for integrating multiple observations made at the same time into a common state estimate. Starting from the formulation of the multi-sensor Kalman filter algorithm, employing a single model for a group of sensors, a set of recursive equations for integrating individual sensor observations can be derived. This method is called "inverse covariance form".

The systems considered to this point are all 'centralized'; the observations made by sensors are reported back to a central processing unit in a raw form where they are processed by a single algorithm in much the same way as single sensor systems. It is also possible to formulate the multi-sensor estimation problem in terms of a number of local sensor filters, each generating state estimates, which are subsequently communicated in processed form back to a central fusion center. This distributed processing structure has a number of advantages in terms of modularity of the resulting architecture. However, the algorithms required to fuse estimate or track information at the central site can be quite complex. This Fusion method is called "track-to-track fusion" and discus as a fourth approach.

In the future equations, Indexes indicate the corresponding sensor number. For example, $z_i(k)$ is the observation of $i^{th}$ sensor at the $k^{th}$ time step.

### 2.2.1. The Group-Sensor Method

The simplest way of dealing with a multi-sensor estimation problem is to combine all observations and observation models in to a single composite 'group sensor' and then to deal with the estimation problem using an identical algorithm to that employed in single-sensor systems. The combinatorial observation vector is defined as:

$$z(k) \sqcup \left[z_1^T(k),...,z_s^T(k)\right]^T, \tag{10}$$

and combinatorial observation model is as:

$$H(k) \sqcup \left[H_1^T(k),...,H_s^T(k)\right]^T \tag{11}$$

and

$$w(k) \sqcup \left[w_1^T(k),...,w_s^T(k)\right]^T, \tag{12}$$

and

$$R(k) = E\{w(k)w^T(k)\} = E\{[w_1^T(k),...,w_s^T(k)]^T [w_1^T(k),...,w_s^T(k)]\} = \\ = blockdiag\{R_1(k),...,R_s(k)\}, \quad (13)$$

Observation noise covariance is as a block-diagonal matrix which each block of diagonal is the observation noise matrix of each sensor. Observation vector are made in a combinatorial manner. Predication equations are like single sensor Kalman filter (equations (1) and (2)).

**2.2.2 The Sequential-Sensor Method**

A second approach to the multi-sensor estimation problem is to consider each sensor observation as an independent, sequential update to the state estimate and for each observation to compute an appropriate prediction and gain matrix. The final formulation provide below:

$$\hat{x}(k|k-1) = F(k)\hat{x}(k-1|k-1) \quad (14)$$

$$P(k|k-1) = F(k)P(k-1|k-1)F^T(k) + G(k)Q(k)G^T(k). \quad (15)$$

$$S_p(k) = H_p(k)P(k|k,p-1)H_p^T(k) + R_p(k) \quad (16)$$

$$W_p(k) = P(k|k,p-1)H_p^T(k)S_p^{-1}(k) \quad (17)$$

$$\hat{x}(k|k) = \left[\prod_{i=1}^{S}(1-W_i(k)H_i(k))\right]\hat{x}(k|k-1) + \sum_{i=1}^{S}\left[\prod_{j=i+1}^{S}(1-W_j(k)H_j(k))\right]W_i(k)z_i(k) \quad (18)$$

For the further information refer to [1].

**2.2.3. The Inverse Covariance Method**

This method was developed to exploit direct equations for filter response. The matrices which are inverted are not so massive. State predication and covariance matrices are as in (1) and (2):

$$P(k|k) = \left[P^{-1}(k|k-1) + \sum_{i=1}^{S}H_i^T(k)R_i^{-1}(k)H_i(k)\right]^{-1} \quad (19)$$

$$\hat{x}(k|k) = P(k|k)[P^{-1}(k|k-1)\hat{x}(k|k-1) + \sum_{i=1}^{S}H_i^T(k)R_i^{-1}(k)z_i(k)] \quad (20)$$

**2.2.4. The Track-to-track Fusion Method**

Track-to-track fusion is an algorithm which combines the estimations which are made at the place of sensors. Indeed, Kalman Filter estimations are made aside for each sensor.

For each single sensor Kalman filter we have:

$$\hat{x}_i(k|k) = \hat{x}_i(k|k-1) + W_i(k)[z_i(k) - H_i(k)\hat{x}_i(k|k-1)], \quad (21)$$

and

$$P_i(k|k) = P_i(k|k-1) - W_i(k)S_i(k)W_i^T(k) \quad (22)$$

which

$$S_i(k) = \left[H_i(k)P_i(k|k-1)H_i^T(k) + R_i(k)\right]. \quad (23)$$

Predictions of local states are made from common states model.

$$\hat{x}_i(k \mid k-1) = F(k)\hat{x}_i(k-1 \mid k-1) \tag{24}$$

and

$$P_i(k \mid k-1) = F(k)P_i(k-1 \mid k-1)F^T(k) + G(k)Q(k)G^T(k) \tag{25}$$

So the path fusion algorithm simply computes a weighted average of paths based on variance weights.

$$\hat{x}_T(k \mid k) = P_T(k \mid k)\sum_{i=1}^{N} P_i^{-1}(k \mid k)\hat{x}_i(k \mid k) \tag{26}$$

$$P_T(k \mid k) = \left[\sum_{i=1}^{N} P_i^{-1}(k \mid k)\right]^{-1}. \tag{27}$$

This method is not an optimal estimation because of correlation between tracks [8], but it is simple and functional.

## 3. PERFORMANCE EVALUATION

### 3.1 Evaluation Theorize

As stated before, computational complexity of fusion algorithms is among the most important factors for its implementation. With respect to matrix operations and its complexity especially for inverse matrix computation, it should be considered as a key factor while hardware or software implementation. More computational load means more powerful and more expensive hardware. From another perspective it needs more time to execute computations.

One of the main factors which impact the computational load is the number of sensors or data fusions sources. This is not necessarily the same for different methods, as there may be a different method for distinct number of sensors which has the least computational load.

In [1], it is mentioned that for inverse matrix calculation, the computational load is proportional to the square of the matrix dimensions. Among the proposed methods, the group sensor method has the most computational complexity. The dimensions of the innovation matrix are proportional to the number of sensors. This matrix should be inverted in each time steps. As a result with the increasing number of sensors, the computational load increases more rapidly.

In sequential-sensor method although the innovation matrix dimensions do not change with the number of sensors, for each sensor an inversing matrix operation has been added. So again the computational load would be increased, but with a linear rate.

Regardless of the number of sensors employed, the largest matrix inversion that is required is of dimension the state vector. The addition of new sensors simply requires that the new of terms $H_i^T(k)R_i^{-1}(k)z_i(k)$ and $H_i^T(k)R_i^{-1}(k)H_i(k)$. Thus the complexity of the update algorithm grows only linearly with the number of sensors employed. In addition, the update stage can take place in one single step.

Excellence of the inverse covariance estimator is more obvious when significant number of sensors is employed. From (10), it is clear that in each cycle of filter, both the prediction covariance matrix and the updated inverse covariance matrix must be inverted. As a result, inverse covariance filter shows its superior characteristics just when the dimensions of the combined observation vector are approximately more than two times of the common state vector dimensions.

## 3.2. Simulation

To proof the claims about the computational load of each method, MATLAB simulation was applied to them in order to evaluate the required process time of each algorithm. In MATLAB, "Profile" function allows the user to measure the required process time for each part of the program.

```
...
Profile on;
...

%codes must be writen here

...
Profile viewer;
...
```

To increase the accuracy of this function, just one of the cores of the processors in operating system was used. Also, the priority of MATLAB software was chosen to be in real time state.

To do this, a target trajectory has been modelled. Then, a noise has been added to this trajectory to model sensors observations. After that, as stated before, Kalman filter multi-sensor data fusion algorithms has been applied to sensors observations to estimate the target track. Simulation results are shown in Fig. 1. Note that our emphasis is on computational load of each method. Also note that just the execution time of the fusion algorithm was measured and path modelling, noise observations and initial values of parameters processes are not included in the mentioned time.

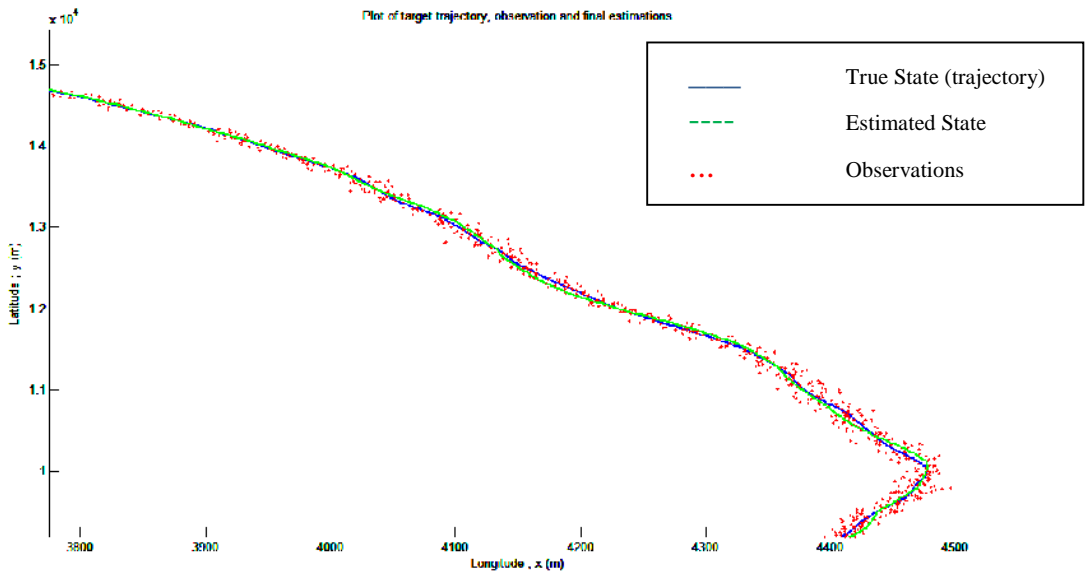

Figure1. True state (trajectory), observation and estimation data of 3 sensors

## 4. SIMULATION RESULTS

The modelled system is simulated and the required process times for each algorithm are listed and shown in TABLE 1 and Fig. 2, respectively.

TABLE 1: The time needed to compute the estimation as a function of increasing number of sensors for each method.

| number of sensors | 1 | 15 | 30 | 45 | 60 | 75 | 90 | 105 | 120 | 150 | 210 | 300 |
|---|---|---|---|---|---|---|---|---|---|---|---|---|
| group sensor | 0.11 | 0.31 | 0.75 | 1.61 | 3.01 | 5 | 7.64 | 11.25 | 15.49 | 30.89 | 75 | 188 |
| sequential sensor | 0.15 | 1.41 | 2.82 | 4.25 | 5.68 | 7.21 | 8.79 | 10.45 | 12.1 | 15.59 | 23.15 | 35.63 |
| invers covariance | 0.2 | 0.37 | 0.56 | 0.74 | 0.92 | 1.1 | 1.29 | 1.46 | 1.66 | 2.03 | 2.74 | 3.86 |
| track to track fusion | 0.1 | 1.14 | 2.71 | 4.08 | 5.42 | 6.76 | 8.12 | 9.45 | 10.8 | 13.51 | 18.9 | 26.94 |

As is seen, in group-sensor method, although a few number of sensors do not need a long time for data fusion, but, as the number of sensors increases, the computations become more time-consuming. Sequential sensor method and track-to-track fusion methods need somewhat the same time. However, these two methods, for the number of sensors greater than 100, are faster than the group-sensor method. For inverse covariance form the story is somewhat different. For the number of sensors less than 20, the computational load is approximately the same as the other methods. However, as the number of sensors increases, the process time is too few in comparison with the others. This is in agreement with what was said in previous sections.

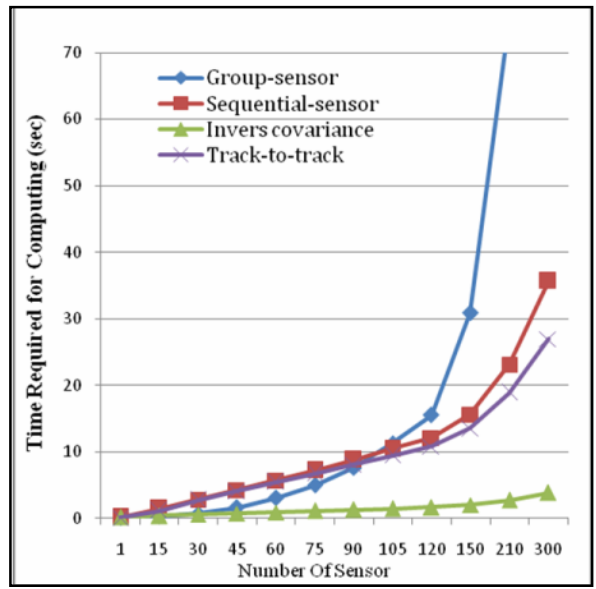

Figure 2. The required time to compute the estimations. It shows the computational complexity of each algorithm.

## 5. CONCLUSION

In order to investigate Multi-sensor single target tracking algorithms base on Kalman filter, four methods were simulated and compared. For this purpose, the time needed to compute the estimation as a function of increasing number of sensors for each method is calculated. In summary, for the number of sensors less than 20, the group-sensor method has the least computational complexity and the inverse covariance method has the second priority. For greater number of sensors, inverse covariance method strongly needs the least process time. Nevertheless, for hardware implementation, type of the hardware (FPGAs, DSPs or so on) and the mathematical relations of algorithms should be taken into account to choose the best method.

## REFERENCES

[1] Hugh Durrant-Whyte, Multi Sensor Data Fusion. Australian Centre for Field Robotics, The University of Sydney NSW 2006.

[2] Y. Bar-Shalom and X.-R. Li, Multitarget-Multisensor Tracking: Principles and Techniques. ISBN 0-9648312-0-1, 1995.B.D.O. Anderson and J.B. Moore. Optimal Filtering. Prentice Hall, 1979.

[3] Y. Bar-Shalom and T. E. Fortmann, Tracking and Data Association. Academic Press, Orlando, FL, 1988.

[4] Jitendra R. Raol, Multi-Sensor Data Fusion with MATLAB, CRC Press, Taylor & Francis Group, 2010.

[5] Zou, Wei, and Wei Sun., " A multi-dimensional data association algorithm for multi-sensor fusion," *Intelligent Science and Intelligent Data Engineering*. Springer Berlin Heidelberg, 2013. 280-288.

[6] Cho, Taehwan, Changho Lee, and Sangbang Choi, "Multi-sensor fusion with interacting multiple model filter for improved aircraft position accuracy," *Sensors* 13.4 (2013): 4122-4137.

[7] Brookner, Eli. Tracking and Kalman filtering made easy. New York: Wiley, 1998.

[8] Y. Bar-Shalom, "On the track to track correlation problem,". *IEEE Trans. Automatic Control*, 25(8):802–807, 1981.